\documentclass[twoside, epsfig]{article}

\input oejv.sty

\begin{document}

\OEJVhead{December 2022}
\OEJVtitle{A study of a contact binary system NSVS 2983201}
\OEJVauth{Debski, B.$^{1}$ and Walczak, K.$^2$} 
\OEJVinst{Astronomical Observatory, Jagiellonian University, Orla 171, 30-244 Kraków, Poland {\tt \href{mailto:b.debski@oa.uj.edu.pll}{b.debski@oa.uj.edu.pl}}}
\OEJVinst{Pabianicka 139C, 95-070 Aleksandrów Łódzki, Poland}

\OEJVabstract{
Here we present the observations and the first light curve analysis of the short-period variable star NSVS 2983201. Using the light curve numerical modeling we find the best fitting model to be of shallow (ff=10\%) contact binary configuration of mass ratio q=0.36. The light curve of the system experiences the O'Connell effect, which led to identifying a large circumpolar starspot. With a careful multi-cases analysis we search for the physical parameters of the system. We find the results obtained with the different methods to be close, but not overlapping. This system will be scheduled for the further monitoring.
}

\begintext

\section{Introduction}\label{secintro} 

The canonical model of the W UMa-type contact binary stars states that these are binary systems consisting of two main-sequence stars sharing a common convective envelope \citep{1968ApJ...151.1123L}. One of the consequences of such a configuration is an almost-constant surface temperature across the entire system. This, in turn, results in a light curve having a primary and secondary minima of a nearly equal depths \citep{1968ApJ...153..877L}, even if the mass ratio of the binary' components is far from $q=M_2/M_1=1$ (where $M_i$ denotes a mass of the i-th component). It is tempting to classify a binary system as a contact binary basing just on the shape of its light curve, and in fact it has been done so in some works, such as \citet{catalina}. This approach doesn't take into account that similarly shaped light curves can be produced by a variety of different type of objects, such as near-contact binaries, low-inclination detached systems, pulsating stars, or even cataclysmic binaries in quiescence \citep{j1622}. For that reason, in order to establish a true nature of the object and derive its physical properties a careful case study, backed with the light curve numerical modeling is needed \citep{debski22}. In this work we are focusing on one of such object, originally suspected as a contact binary star. We begin this work by outlying the catalog of the suspected contact binaries (Sec.~\ref{sec:catalog}), then we present the object (Sec.~\ref{sec:object}) and the observational setup (Sec.~\ref{sec:obs}). The methods of the analysis and the numerical modeling is described in Sec.~\ref{sec:data}. The calculation of the physical properties of the studied object is described in Sec.~\ref{sec:param} along with the results. We finish with a short summary in Sec.~\ref{sec:summary}.

\subsection{The Catalog}
\label{sec:catalog}
We have compiled a list of poorly studied objects suspected to be contact binaries. The list hosts objects with the EW-type light curve \citep{kopal}, which are suitable for observations from the latitude c. 50$^{\circ}$\,N. The list will be henceforth referred to as the Krakow EW-type Light curve (KEWL) Catalog\footnote{\url{http://bade.space/ew/}}. It contains the mean magnitude (primarily in the Bessel R or V bands), sky coordinates, orbital period, the total-eclipse flag, time of the last observations taken, the objects' light curve, sky chart (30' x 30'), a link to the discovery source, and a link to the SIMBAD database \citet{simbad}. Each object within KEWL has its internal ID which consists of the prefix 'KR' (Krakow Register) and the five-digit index number.

\subsection{The Object}
\label{sec:object}
Here we present a study of KR00245 (NSVS 2983201, GAIA DR3 2253442557073280000, RA (J2000): 18h 42m 09s, DEC (J2000): $+64^{\circ}$ 13' 46''). This object was listed as a contact binary by \citep{gettel}, but until now it had a very poor light curve coverage and no numerical modeling analysis performed. We chose to observe this object primarily because of its short orbital period ($P = 0.285999$\,d.), brightness, visibility, and the fact that it experiences total eclipses. The latter was determined by the preliminary screening observations we took in June 2020. A total eclipse is the key factor in the process of numerical modeling. It allows to establish the mass ratio and inclination, which otherwise are highly entangled (cf. \citet{2003CoSka..33...38P} and \citet{terrell}). Contact binaries with no total eclipse require their mass ratio to be established via studying the radial velocities. As most contact binaries have no spectroscopically determined mass ratios, we usually limit our studies to the binaries with total eclipses.

\section{The Observational Setup}
\label{sec:obs}
\begin{table}
\caption{Identifiers in GSC 2.4 \citep{gsc} and GAIA DR3 \citep{gaiadr3} and coordinates (J2000) of the variable and the three comparison stars}\vspace{3mm}  
\centering
\begin{tabular}{lcccc}
\hline
    Star    &   GSC ID          &   GAIA ID         & RA [h m s]    & DEC [$^{\circ}$ $'$ $''$] \\ \hline \hline
    Var     &   GSC 04223-00037 &   Gaia DR3 2253442557073280000 &   18 42 09.5  &   +64 13 46.6  \\
    Comp1   &   GSC 04223-00717 &   Gaia DR3 2256444842292998272 &   18 41 10.4  &   +64 16 51.1  \\
    Comp2   &   GSC 04223-00549 &   Gaia DR3 2253443381707006720 &   18 41 58.7  &   +64 17 21.4  \\
	Comp3   &   GSC 04223-00685	&   Gaia DR3 2253443759664125952 &   18 41 48.6  &   +64 17 31.3  \\
\hline\end{tabular}\label{tab:coord}
\end{table}

\begin{table}
\caption{Mean magnitudes of the variable and the three comparison stars, taken from \citet{gaiaspectra}. The effective temperatures are taken from \citet{gaiadr3}. We chose the Comp1 as the primary comparison star because of the similarity of its effective temperature to the one of the variable star.}\vspace{3mm}  
\centering
\begin{tabular}{lcccccc}
\hline
    Star    & $B$ [mag]    & $V$ [mag]  & $R$ [mag]  & $I$ [mag]  & $T_{eff}$ [K] \\ \hline \hline
    Var     &   14.114(22) & 13.172(13) & 12.809(13) & 12.465(16) &  5512 \\
    Comp1   &   13.920(3)  & 13.207(2)  & 12.803(2)  & 12.412(1)  &  5517 \\
    Comp2   &   14.167(3)  & 13.621(2)  & 13.314(2)  & 13.031(2)  &  6230 \\
	Comp3   &   14.490(3)  & 13.806(2)  & 13.427(2)  & 13.080(1)  &  5610 \\
\hline
\end{tabular}\label{tab:bvri}
\end{table}

\begin{figure}[ht]
\centering
\includegraphics[width=10cm]{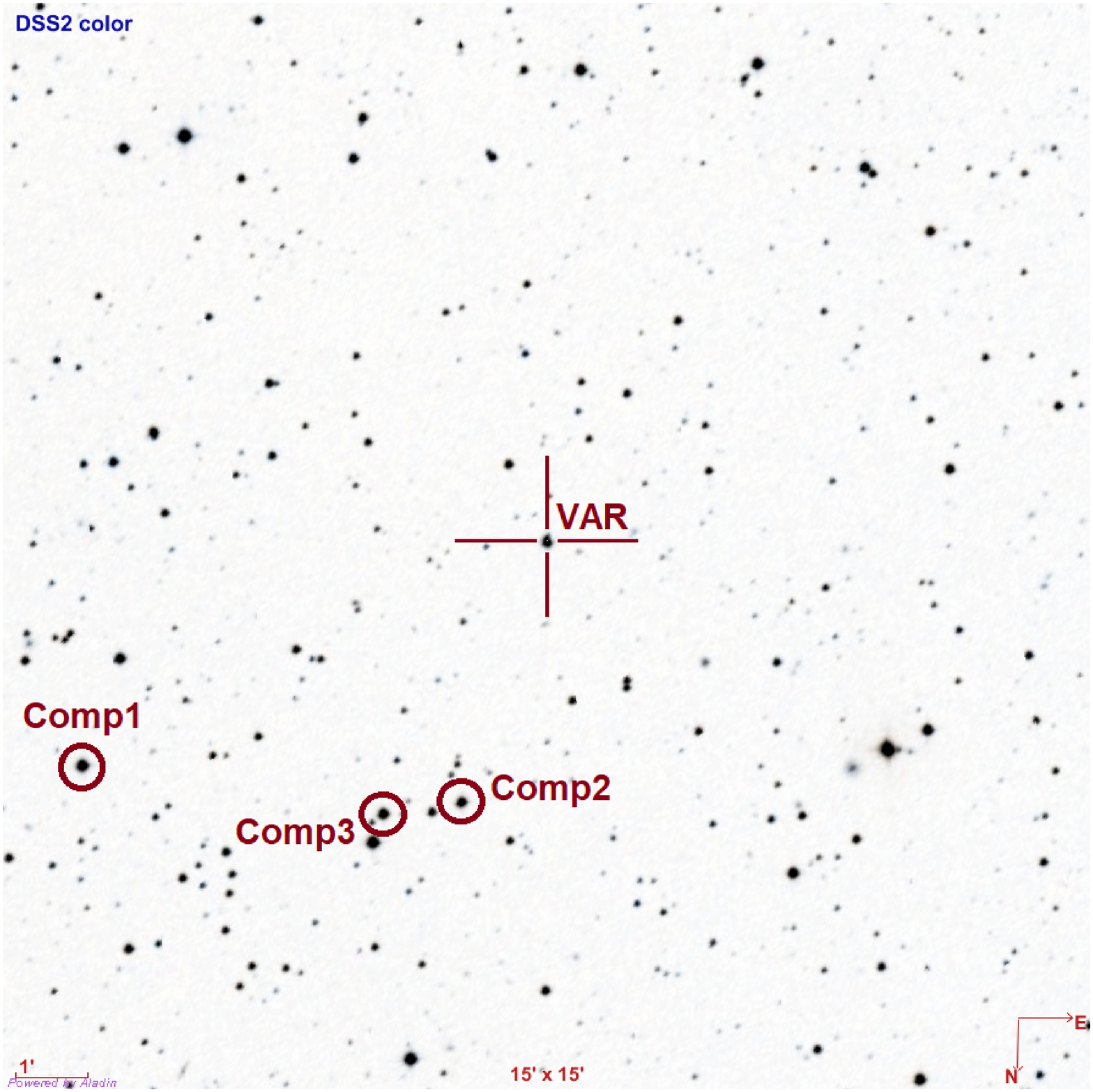} 
\caption{The chart of sky around KR00245. The chart has 15' by 15' size, and is oriented North-down, East-right. The FOV of our CCD camera was shifted slightly to the East, so that the primary comparison star wouldn't be so close to the edge of the frame.}
\label{fig:chart}
\end{figure}

KR00245 was observed as a Target of Opportunity with the Cassegrain telescope (aperture 500 mm, effective focal length 6650 mm) in the Astronomical Observatory of the Jagiellonian University in Krakow on the night of 07/08.10.2022. The photometry was performed with the Apogee Alta U42 camera in the D9 casing. The CCD chip maintained a stable temperature of $-30^{\circ}$\,C throughout the work. The observations started at JD\,=\,2459860.237 and lasted continuously for 0.303\,d, covering the entire orbital period. We took 190 photometric frames in each filter: B, V, R and I. The exposure times were: 40\,s, 30\,s, 20\,s, and 20\,s, respectively. The calibration frames were collected on the same day (the on-sky Flatframes in each filter in the evening, while Bias- and Darkframes directly after the observations).

Within the field view of the camera (15' by 15') we found three objects suitable to serve as comparison stars. For them, as well as for KR00245 we extracted the brightness in the B, V, R, and I filters using the synthetic photometry from GAIA Mission low-resolution spectra \citep{gaiaspectra}. Because the brightness and color of the variable star is build upon the averaged G$_{\rm B}$ and G$_{\rm R}$ bandpasses, we decided to use the GAIA effective temperatures as a more reliable indicator for choosing the best comparison star. We settled on using the Comp1 (GSC 04223-00717, see the sky chart in Figure~\ref{fig:chart}) as the comparison star for this study. The designations and coordinates of the all four objects are in Table~\ref{tab:coord}. Their B, V, R, and I brightnesses, and the effective temperatures are in Table~\ref{tab:bvri}. 

\section{Data analysis and the modeling results}\label{sec:data}

\subsection{The photometry}
In this work we performed a standard Bias-, Dark-, and Flatframe calibration using the C-Munipack v2.1.33 astronomical software\footnote{\url{https://c-munipack.sourceforge.net}} offering the DAOPHOT-based aperture photometry. An average size of a stellar object in our data is FWHM\,$\approx 2.3$\,px, which translates to about FWHM\,$\approx 1.8$\,arcsec, a rather typical value for the Kraków site.

We determined the moment of the primary minimum with the Kwee \& van Woerden method \citep{kwee} at:
\begin{equation}
    HJD_{Min_I} = 2459860.468679(191)
\end{equation}

In this work we define the primary minimum as the one during which the less massive component of the binary is eclipsing the more massive one. The light curve we gathered is burdened with the O'Connell effect, i.e. there is a difference between the height of the brightness maxima. In fact, this asymmetricity weights on the entire phased light curve, skewing the primary minimum slightly to the left. Because the Kwee \& van Woerden method assumes symmetrical profile of the minimum, we expect that the determined here $HJD_{Min_I}$ might not fall exactly to the middle-of-an-eclipse. 

\subsection{Light curve numerical modeling}

\begin{table} [t]
\caption{The parameters of the best fitting model resulting from the light curve numerical modeling. The errors are adopted at the 2-$\sigma$ confidence level. The right-hand side of the table presents the fill-out factor and the fractional radii of the components, calculated from the $\Omega$ and $q$ of the best fitting model from the left-hand side of the table.}\vspace{3mm}  
\centering
\begin{tabular}{rlcrl}
\hline
\multicolumn{2}{c}{Best fitting numerical model parameters} &\,\hspace{0.5cm}\,& \multicolumn{2}{c}{Geometry parameters} \\
	 Parameter & Value(error) & & Parameter & Value(error)          \\ \hline \hline
\,$i$   					& $87.8(1.0)^{\circ}     $ && \,$ff$  			  & $9.5(4.6)\% $ \\
\,$T_1$ 					& $5512\,{\rm K}$\,[fixed] && \,$r_{1,{\rm side}}$ & $0.4778(27) $ \\ 
\,$T_2$ 					& $5685(26)\,{\rm K}     $ && \,$r_{2,{\rm side}}$ & $0.2908(26) $ \\  
\,$\Omega$ 					& $2.571(11) 		     $ && \,$r_{1,{\rm back}}$ & $0.5054(34) $ \\ 
\,$q$, $\frac{m_2}{m_1}$    & $0.3585(47)		     $ && \,$r_{2,{\rm back}}$ & $0.3266(42) $ \\
\,$L_{1,{\rm B}}$ 			& $8.253(77) 			 $ && \,$r_{1,{\rm pole}}$ & $0.4457(21) $ \\  
\,$L_{2,{\rm B}}$  			& $3.822(58) 			 $ && \,$r_{2,{\rm pole}}$ & $0.2785(22) $ \\  
\,$L_{1,{\rm V}}$  			& $8.361(72) 			 $ && \,$r_{1,{\rm geom}}$ & $0.4757(28) $ \\  
\,$L_{2,{\rm V}}$  			& $3.779(49) 			 $ && \,$r_{2,{\rm geom}}$ & $0.2979(29) $ \\  
\,$L_{1,{\rm R}}$  			& $8.505(65) 			 $ &&&\\   
\,$L_{2,{\rm R}}$  			& $3.759(43) 			 $ &&&\\   
\,$L_{1,{\rm I}}$  			& $8.565(61) 			 $ &&&\\   
\,$L_{2,{\rm I}}$  			& $3.724(39) 			 $ &&&\\   
\,Spot co-latitude 			& $9.8(3.3)^{\circ}      $ &&&\\   
\,Spot longitude 			& $142.1(6.7)^{\circ}    $ &&&\\   
\,Spot radius 				& $35.7(3.3)^{\circ}     $ &&&\\   
\,Spot temp  				& $0.75\,T_1$\,[fixed]	   &&&\\ 
\hline
\end{tabular}\label{tab:model}
\end{table}

We subjected the phased light curve to the numerical modeling. For this process we used a modified Wilson-Devinney code \citep{wilson} with the Price controlled Monte Carlo \citep{zola97} search method put in place of the differential corrections mechanism. This program was used e.g. in \citet{zola10}, \citet{debski20} or \citet{debski22}. For the numerical simulations we used a light curve generator based on the same version of the modified Wilson-Devinney code. During the modeling we are assuming a square root limb darkening law with the limb darkening coefficients taken from \citet{claret11} and \citet{claret13}. The gravitational brightening coefficient was fixed to $\beta=0.08$ (g$\,=0.32$) \citep[][]{1968ApJ...153..877L} and the albedo was fixed at $A=0.5$ \citep[][]{1969AcA....19..245R}. The distortion of the light curve led us to believe that this particular object experiences a presence of a starspot. Because of that, we a circular spot parametrized by its radius, temperature, and the co-latitude and longitude of its center. The spot is assumed to be of magnetic origin hence visible in the visual wavelengths as a subluminuous region on the surface of a star \citep{mullan}. Because the primary star would have a thicker convective envelope, we chose it as a spot host.

\begin{figure}[t!]
\centering
\includegraphics[width=0.9\linewidth]{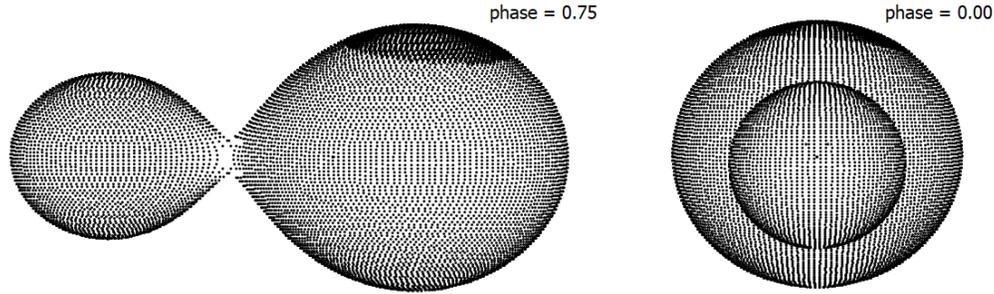} 
\caption{A graphical depiction of the best fitting model of KR00245, taken ad the orbital phases $\phi=0.75$ and $\phi=0.0$.}
\label{fig:model}
\end{figure}

\begin{figure}[t!]
\centering
\includegraphics[width=0.9\linewidth]{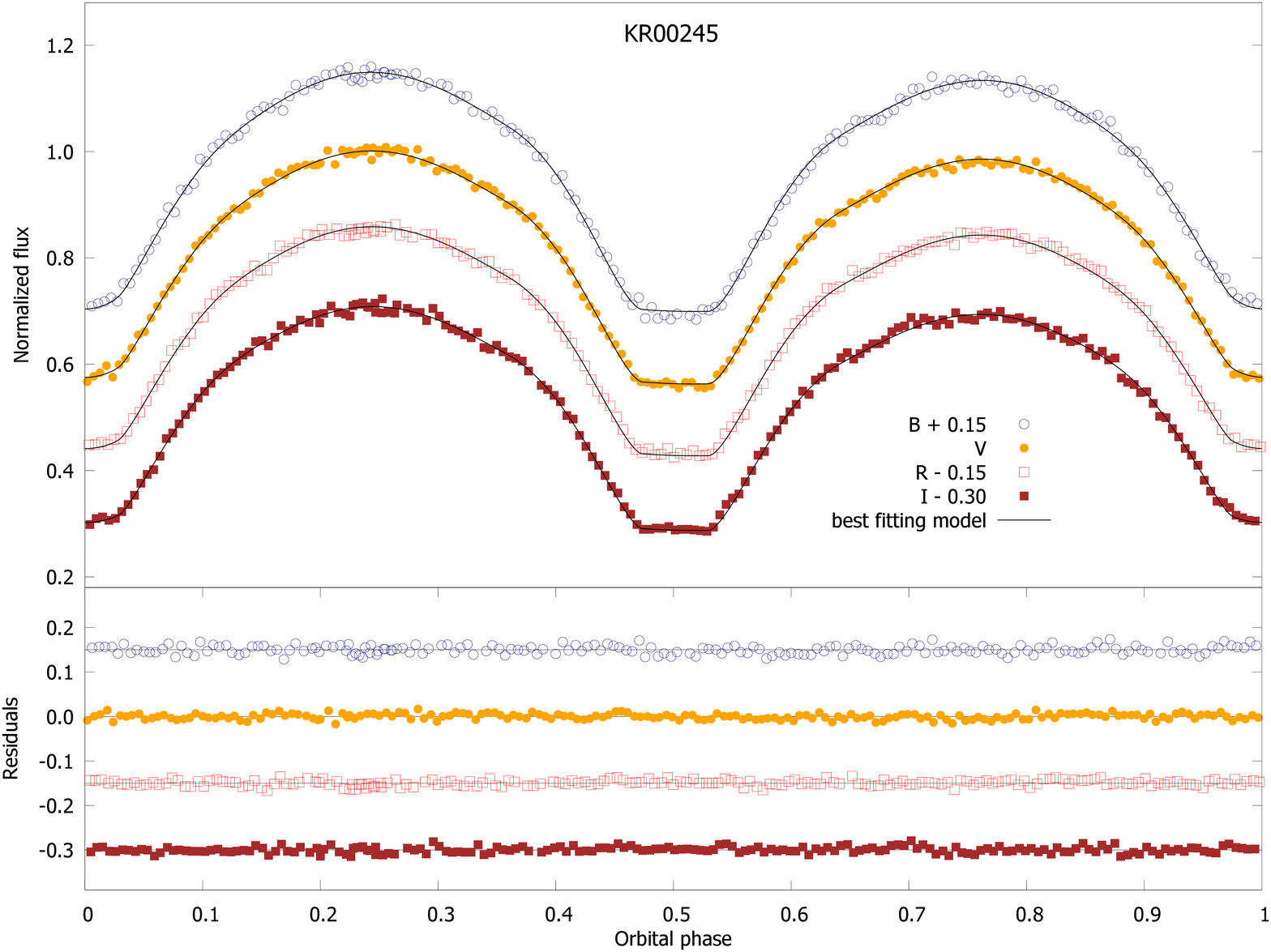} 
\caption{The four phased light curves: in B, V, R, and I filters, superimposed with the synthetic light curves build on the best fitting model. The the light curves are in the flux units, normalized to unity at the orbital phase $\phi=0.25$. The bottom panel shows the residuals left after subtracting the synthetic light curve from the observations. The vertical shifts are introduced for the clarity of the picture.}
\label{fig:lc}
\end{figure}

The best fitting model is presented na Table~\ref{tab:model} along with the 2-$\sigma$ uncertainties. We calculated the geometrical parameters: the fill-out factor, $ff$ and the fractional radii, $r_i,j$ (where $i$ stands for the number of the binary component and, $j$ for the type of radii: $side$, $back$, $pole$ and the geometrical, $geom$.) The former is defined here as following:
\begin{equation}
    ff = \frac{\Omega_{L_1}-\Omega}{\Omega_{L_1}-\Omega_{L_2}},
\end{equation}

\noindent where $\Omega_{L_1}$ is the Roche pseudopotential at the inner critical Lagrange surface, $\Omega_{L_2}$ - analogously at the outer critical Lagrange surface, and $\Omega$ is the pseudopotential of the system (i.e. the surface of the binary). The geometrical parameters were calculated using the Roche Geometry Calculator available at our project's website\footnote{\url{http://bade.space/soft/}}. The graphical depiction of the best fitting model is presented in Figure~\ref{fig:model}. This was prepared using the picture generator based on the newest version of the W-D code \citep{2020ascl.soft04004W} accessible via the interface on the aforementioned web page. In the Figure~\ref{fig:lc} we show the observed light curves in the B, V, R and I filters, superimposed with the synthetic light curves presenting the best fitting model from Table~\ref{tab:model}.

\section{The physical parameters}
\label{sec:param}
We calculate the individual masses of the binary' components, $M_1$ and $M_2$ by solving for the total system luminosity, $L_T$ first. To calculate the total luminosity we must find the absolute magnitude of the system first. This will be done using two independent methods.

\subsection{Absolute magnitude and the distance calculation}

The first approach involves calculating the absolute magnitude in the V filter, $M_{\rm V}$, from the distance, $D$, to the object. By employing the GAIA Mission parallax of KR00245, $\pi = 2.498(45)$, the GAIA-based distance is:
\begin{equation}
    D_{G} = 1304(8)\,ly.
\end{equation}
\noindent Then, by definition the GAIA-based absolute magnitude is:
\begin{equation}
M_{\rm V,G} = m_{\rm V} - A_{\rm V} - 5{\rm log}D + 5 = 4.713(28),
\end{equation}

\noindent where $A_{\rm V} = 0.0147(2)$ is the interstellar extinction taken from \citet{schlafly} using the IRSA DUST web application\footnote{\url{https://irsa.ipac.caltech.edu/applications/DUST} (accessed November 2022)}.

The second approach to find the absolute magnitude is to use the \citet{rucinski} relation:
\begin{equation}
    M_{\rm V,R} = -4.44{\rm log}P + 3.02 (B-V)_0 + 0.12,
\end{equation}
\noindent where the $M_{\rm V,R}$ stands for the 'Rucinski-based' absolute magnitude, and $(B-V)_0$ denotes the dereddened color of the system. If we take the color of the system directly from GAIA measurements (i.e. from Table~\ref{tab:bvri}):
\begin{equation}
(B-V)_{G} = 0.942(26),
\end{equation}
\noindent and we take into account the color excess: (based on the $A_V=3.1\,E(B-V)$ relation):
\begin{equation}
E(B-V) = 3.1A_V = 0.047(1),
\end{equation}
\noindent then the dereddened GAIA-based color is:
\begin{equation}
    (B-V)_{0,G} = (B-V)_{G} - E(B-V) = 0.8947(27).
\end{equation}
\noindent Applying the above to the Rucinski' formula, we get the absolute magnitude:
\begin{equation}
M_{V,R} = 5.236(78).
\end{equation}
The distance to the object calculated via the distance modulus is then:
\begin{equation}
D_{R,G} = 10^{1+0.2(m_v-M_V-A_V)} = 1026(34) \,ly.
\end{equation}
\noindent The distance and the absolute magnitude differs significantly form the one found with the parallax method. Here we stumble upon a possible caveat. We are reasoning that the color of the objects based on the GAIA synthetic photometry catalog (i.e. as in the Table~\ref{tab:bvri}) works very well for constant stars (or at least for stars with a slow brightness variation). Because our contact binary experiences a tentative spot phenomenon, the light curve intrinsic variability might produce an uneven statistical distribution in GAIA photometry. Hence, the colors might be somewhat affected, and less reliable.

If we compute the colors basing on our observations, and we take Comp1 and the reference color star, then the color of KR00245 is:
\begin{equation}
    (B-V)_{obs} = 0.718(16),
\end{equation}
\noindent and the dereddened color is:
\begin{equation}
    (B-V)_{0,obs} = 0.6717(17).
\end{equation}
\noindent Using again the Rucinski's formula we get the observation-based absolute magnitude in V filter:
\begin{equation}
    M_{V,obs} = 4.562(48),
\end{equation}
\noindent and the observation-based distance:
\begin{equation}
   D_{R,obs} = 1399(31)\,ly.
\end{equation}
\noindent These are in much better agreement with the distance and absolute magnitude found with the parallax method.

Finally, we can calculate the expected color of the system using the linear period-color relation for contact binaries with orbital period $P<0.45$\,d from \citet{debski22}:
\begin{equation}
     (B-V)_{0,D} = (0.905\,\pm\,0.138) - (0.847\,\pm\,0.374)P = 0.663\,\pm\,0.245.
\end{equation}
\noindent Athough the uncertainty of the latter is rather high, this is still in a fine agreement with the GAIA- (parallax-) and observation-based system dereddened colors. For the remaining of this paper we will be using the GAIA- and the observation-based absolute magnitudes.
    
\subsection{Masses calculation}

To calculate the masses, we first find the total luminosity of the system using the formula:
\begin{equation}
\label{eq:lum}
    L_{T}\, [L_{\odot}]=10^{-0.4(M_{\rm V}-4.83)}\,.
\end{equation} 
In the next step we calculate the orbital separation. To do that, we use the fact the orbital separation is the scaling factor for the fractional radii. Incorporating this into the Stefan–Boltzmann law we obtain the relation:
\begin{equation}
\label{eq:separ}
    A \,[R_{\odot}] = \sqrt{\frac{L_{T}}{T_1^4r_1^2+T_2^4r_2^2}}.
\end{equation}
Then, from the Kepler's Third Law we get the total system mass expressed in Solar masses:
\begin{equation}
\label{eq:mtot}
    M_{tot} \,[M_{\odot}]=\frac{1}{74.53}\frac{A^3}{P^2}
\end{equation}
To calculate the individual component masses we combine the total system with the mass ratio $q$ found in the process of the numerical modeling:
\begin{equation}
M_1 = \frac{M_{tot}}{1+q},
\end{equation}
\begin{equation}
M_2 = \frac{M_{tot}}{1+\frac{1}{q}}.
\end{equation}

We calculated the individual masses, as well as the values of all of the intermediate variables for two cases: the GAIA-based absolute magnitude, $M_{\rm V,G}$, and the observation-based absolute magnitude, $M_{V,obs}$. The final results are stored in Table~\ref{tab:kr245}

\begin{table}
\caption{The final set of the physical parameters of KR00245. All system parameters: total system luminosity, $L_{T}$, the orbital separation, $A$, and the individual masses of the components: $M_1$ and $M_2$ are calculated twice, each for the absolute magnitude, $M_{\rm V}$, obtained with a different method. The first row, GAIA+OBS, contains the values based on the GAIA Mission parallax and the modeling results. The second row, RUC+OBS, contains the solution based on the Rucinski's absolute magnitude formula with the system color adopted from our observations, and the modeling results.}\vspace{3mm}  
\centering
\begin{tabular}{rcccccc}
\hline
            & $D$ [ly]  & $M_{\rm V}$ & $L_{T}$ [L$_{\odot}$] & $A$ & $M_1$ & $M_2$ \\ \hline \hline
   GAIA+OBS &  1304(8)  &  4.713(28)  & 1.114(28) & 2.044(48) & 1.031(74) & 0.370(27) \\
   RUC+OBS  &  1399(31) &  4.562(48)  & 1.280(56) & 2.191(74) & 1.270(132) & 0.456(47) \\ 
\hline
\end{tabular}\label{tab:kr245}
\end{table}

\section{Conclusions or Summary}\label{sec:summary}

We have shown in this work the first case study of the NSVS 2983201 using our original B, V, R, I photometry. By the means of the light curve numerical modeling we have shown this object best fitting model is of a shallow contact binary configuration with mass ratio of $q=0.3585(47)$. Using two independent methods, we established the absolute magnitude of the system. Basing on both cases of the established absolute magnitude we calculated the physical parameters. The only exception are the stellar radii, which were calculated by us earlier, in Table~\ref{tab:model}, expressed in the units of the orbital separation, $A$. The calculation of the physical parameters with the two different absolute magnitudes resulted in obtaining very close, yet significantly differing values. This can be explained partially by the presence of a large cool spot on the primary component of the binary. 

We acknowledge there is a blended light source very close to our object, partially falling within the boundary of our aperture. We did not take this source into account during the modeling, since it has a negligible brightness contribution to our photometry (less than 0.5\%).

NSVS 2983201 a.k.a. KR00245 will be subjected to further studies inclined towards the search for the spot migration and its influence on the system parameters. We infer the spot is migrating (or {\it evolving}, broadly speaking) after comparing the light curve parameters from \citet{gettel} and our observations. In their work the V-band amplitude of the light curve variation is $Amp = 0.674$\,mag, while in our observations is is about $Amp = 0.608(4)$. In addition, the short orbital period of the binary suggest this object should experience the spot-migration-induced intrinsic light curve variability, as shown in \citet{debski22}. The linear Period-Color relation for short-period contact binaries from therein returns the dereddened color of the binary very close to the one found in this work.

\end{document}